\documentclass[conference]{IEEEtran}
\IEEEoverridecommandlockouts
\usepackage[cmex10]{amsmath}
\usepackage{amssymb}
\usepackage{amsthm}
\usepackage{subfigure}
\usepackage{amsthm,bm}
\usepackage{balance}
\usepackage{soul}
\usepackage{mhchem}
\allowdisplaybreaks

\usepackage{graphicx,epsf,color,url,booktabs,tabularx}
\usepackage[linesnumbered,ruled,vlined]{algorithm2e}
\usepackage[caption=false,font=normalsize,labelfont=sf,textfont=sf]{subfig}
\usepackage{diagbox}

\def\BibTeX{{\rm B\kern-.05em{\sc i\kern-.025em b}\kern-.08em
    T\kern-.1667em\lower.7ex\hbox{E}\kern-.125emX}}
\IEEEoverridecommandlockouts

\begin{document}

\title{An Ion-Intercalation Memristor for Enabling Full Parallel Writing in Crossbar Networks}

\author{
\IEEEauthorblockN{Tingwei Zhang}
\IEEEauthorblockA{\textit{Oregon State University}\\
Corvallis, OR, USA \\
tingwei.zhang@oregonstate.edu}
\vspace*{-0.22in}
\and
\IEEEauthorblockN{Jiahui Liu}
\IEEEauthorblockA{\textit{Clemson University}\\
Clemson, SC, USA\\
jiahui@clemson.edu}
\vspace*{-0.22in}
\and
\IEEEauthorblockN{David Allstot}
\IEEEauthorblockA{\textit{Carnegie Mellon University}\\
Pittsburgh, PA, USA\\
dallstot@andrew.cmu.edu}
\vspace*{-0.22in}
\and
\IEEEauthorblockN{Huaping Liu}
\IEEEauthorblockA{\textit{Oregon State University}\\
Corvallis, OR, USA \\
huaping.liu@oregonstate.edu}
\vspace*{-0.22in}
}

\maketitle

\begin{abstract}
Crossbar architectures have long been seen as a promising foundation for in-memory computing, using memristor arrays for high-density, energy-efficient analog computation. However, this conventional architecture suffers from a fundamental limitation: the inability to perform parallel write operations due to the sneak path problem. This arises from the structural overlap of read and write paths, forcing sequential or semi-parallel updates and severely limiting scalability. To address this, we introduce a new memristor design that decouples read and write operations at the device level. This design enables orthogonal conductive paths, and employs a reversible ion doping mechanism, inspired by lithium-ion battery principles, to modulate resistance states independently of computation. Fabricated devices exhibit near-ideal memristive characteristics and stable performance under isolated read/write conditions.
\end{abstract}


\vspace*{-0.12in}
\section{Introduction} \label{intro}
Rapid advances in artificial intelligence (AI) in the past decade have driven demand for more efficient, scalable, and energy-conscious computing hardware. Traditional von Neumann architectures, though well-suited for general-purpose computation, face key limitations for data-intensive AI workloads, most notably the memory wall. This arises from the physical separation of memory and processing units, causing latency and energy overhead due to heavy data movement. As machine learning models grow larger and more complex, this bottleneck increasingly limits performance \cite{yao2020fully}.

In-memory computing has emerged as a promising solution to these challenges, with the memristor
at its core that stores data as resistance states and enables analog computation \cite{roy2020memory,bavikadi2020review}.
In particular, memristor-based crossbar arrays provide an energy-efficient platform for matrix-vector
computation, a fundamental operation in deep learning \cite{song2023recent}. By leveraging Ohm's and Kirchhoff's laws, these architectures enable massively parallel analog computation, eliminating the need for separate arithmetic units and memory access cycles, which can account for as much as 90\% of the total energy consumption \cite{Horowitz2014}.

Despite their theoretical advantages, however, memristor-based crossbar systems face persistent engineering challenges that limit their practical deployment. The most prominent is the sneak path problem, where unintended current through unselected devices degrades read and write accuracy \cite{shi2020research}. Various circuit-level (e.g., 1T1R, selector devices) and algorithmic (e.g., sparse updates) solutions have been proposed \cite{lee2025recent,chen2015fault,manem2010design,GUL20191091,cassuto2016write,chen2003nanoscale,9021944,8613449,datta2023impact,9798246,10181867,mi12010050}, but these often come with trade-offs, which ultimately make fully parallel write operations infeasible.

The sneak path problem originates from a fundamental structural issue. By the design of the crossbar network, memristors on the same row share the same voltage rail, which is intended to enable efficient multiply-and-accumulate (MAC) operations during the readout stage. On the other hand, in conventional two-terminal memristor designs, read and write signals share the same physical path. Thus, both readout and write-in stages share the same circuit topology. During the write-in stage, the shared voltage rail inevitably exposes non-target memristors to the same write-in voltage, forming a parallel circuit with multiple unintended current paths around the target cell, the essence of the sneak path problem.
Although approaches like 1T1R or 1S1R introduce partial access control

to suppress sneak currents, they only provide semi-independent switching because one end of the write path remains shared across rows or columns, inherently limiting the scalability of simultaneous multi-cell programming. Consequently, only one switch can be active per row. Thus, for a network with a size of $M \times M$ cells, updating the entire network requires either iterating through each individual cell (${\cal O}(M^2)$), or row-wise parallel (${\cal O}(M)$) \cite{zhang2023edge}.

In this sense, the lack of true parallel writes is not merely a side effect of sneak paths, but a direct consequence of the crossbar circuit structure, specifically the use of shared voltage rails for MAC operations, combined with the two-terminal device configuration that couples read and write paths. This structural insight underscores the need for architectural decoupling of read and write paths to achieve genuinely parallel, scalable configuration.

We propose a memristor that leads to architecture revolution for in-memory computing using memristor arrays. Unlike conventional filament-based memristors that rely on nanoscale defect migration and suffer from stochastic drift, our device uses a statistical resistivity model that leverages ion intercalation into an ion acceptor layer, where the overall ion concentration governs conductance. The bulk statistical property of this modulation ensures high stability and reproducibility; the structure physically separates the read and write paths
to ensure read-write orthogonality, eliminating interference between computational and programming signals and enabling true parallel write capability. Furthermore, it supports a parallel programmable network via digital logic and independent control signals, reducing write time complexity from ${\cal O}(M^2)$ to ${\cal O}(1)$ regardless of array size.

\section{Related Works}
A variety of techniques have been proposed to suppress sneak path interference and improve write reliability in memristor crossbar arrays. Most fall into three categories \cite{lee2025recent}:
\begin{itemize}
    \item Adding access control elements such as transistors or selectors.
    \item Using half-select voltage schemes or serialized time-domain signaling.
    \item Hybrid approaches combining structural gating and temporal scheduling.
\end{itemize}
Here, we briefly review representative techniques from each category,
examining their
tradeoffs and
challenges scalability in high-density or parallel-write scenarios.

\subsection{Control Elements}
Adding access control elements such as transistors, diodes or selectors is
a commonly used approach for sneak path mitigation. The basic idea is
to introduce a switch to the memristor, thus selectively enabling the memristor cell.
In \cite{manem2010design}, Manem {\it et al.} proposed a solution
to address the sneak path issue
by adding a diode to each memory cell, producing a new cell of one diode and one memristor (1D1M).
This scheme decreases the output swing because of the diode threshold voltage.
Gul {\it et al.} fabricated a ZnO-based Schottky diode in a 1D1M configuration,
validating its low-power, anti-crosstalk capabilities for crossbar scaling \cite{GUL20191091}.
Chen {\it et al.} introduced a sneak path mitigation method
by introduced a transistor
in series with the memristor, producing a new cell of one transistor and one memristor (1T1R) \cite{chen2015fault}.
A third voltage rail
is required in this structure to rectify the transistor, allowing write operations.
In another example, Humood {\it et al.} demonstrated that
a one-selector, one-memristor (1S1R) configuration reduces sneak path current by up to $200\times$ and significantly improves read margins and device selectivity, enabling up to $4k\times4k$ crossbar scaling \cite{9021944}.

While 1T1R and 1S1R structures provide effective access gating and sneak path suppression, they do not fundamentally alter the sequential nature of programming: only one device can be reliably written at a time. As such, these structures still rely on polling-based update protocols, with write throughput scaling poorly for large arrays.

\subsection{Half-select Voltage}
Half-select biasing and serialized time-division signaling offer circuit-level mitigation for sneak path interference by carefully controlling the voltage applied to non-targeted rows and columns of the array during memory operations. The basic principle of half-select schemes is to apply a reduced voltage (typically $V/2$ or $V/3$) to all non-selected wordlines and bitlines, while applying the full write voltage only to the selected line pair. This ensures that only the targeted memristor receives the full voltage drop required for switching, while adjacent ``half-selected" devices remain below the threshold for unintended state change.
Chen {\it et al.} proposed adaptive thresholding based on coding-assisted estimation to probabilistically reduce sneak-path-induced read errors \cite{8613449}.
They also explored a bilayer-stacked RRAM architecture where the sneak path current is reduced by 20\% using a voltage-stress-tolerant half-read scheme, improving read margins and reducing degradation under high-density integration scenarios \cite{mi12010050}.

Half-select biasing and time-division techniques offer circuit-level solutions to minimize write disturbance and leakage currents. However, these methods generally require serialized write sequences and precise timing control, limiting the achievable configuration throughput as array size scales.

\subsection{Hybrid Approaches}
A range of opportunistic techniques have
also been explored to suppress sneak paths through unconventional means such as timing control, differential encoding, and peripheral circuit design. These approaches aim to mitigate interference without adding complex selector elements or fundamentally altering the crossbar topology.

Jain {\it et al.} demonstrated a selector-less platform
that employs switched-capacitor sensing and differential weight encoding, enabling energy-efficient parallel programming and inference in neuromorphic applications \cite{10181867}.
In \cite{9798246}, Yang introduced the concept of a ``timing selector" using voltage-dependent transient delays to mitigate sneak paths while preserving linear $I$-$V$ behavior for compute-in-memory applications.
Datta {\it et al.} compared hybrid 1T1R array configurations and highlighted that multi-row structures tolerate sneak paths better in logic synthesis tasks than single-row topologies \cite{datta2023impact}.

These techniques, while innovative, remain constrained by the shared conductive paths in conventional two-terminal arrays. As such, they typically operate within serialized or partially-parallel regimes, failing to offer scalable true-parallel write capabilities for large crossbar arrays.

In summary, existing efforts have contributed significantly to improving the stability of memristor arrays under sneak path conditions.
However, these solutions require sequential configuration as a necessary compromise.
This paper develops a structural decoupling strategy to enable scalable, parallel write operations,
shifting the focus from suppressing interference to redesigning access.
The core of the solution developed here is a new memristor architecture, a four-terminal device, that physically decouples the read and write paths. By structurally isolating data access from resistance programming, the proposed design enables stable, interference-free parallel writing across the entire array.

\section{Methodology}
\begin{figure}[h]
    \centering
    \includegraphics[width=0.95\columnwidth]{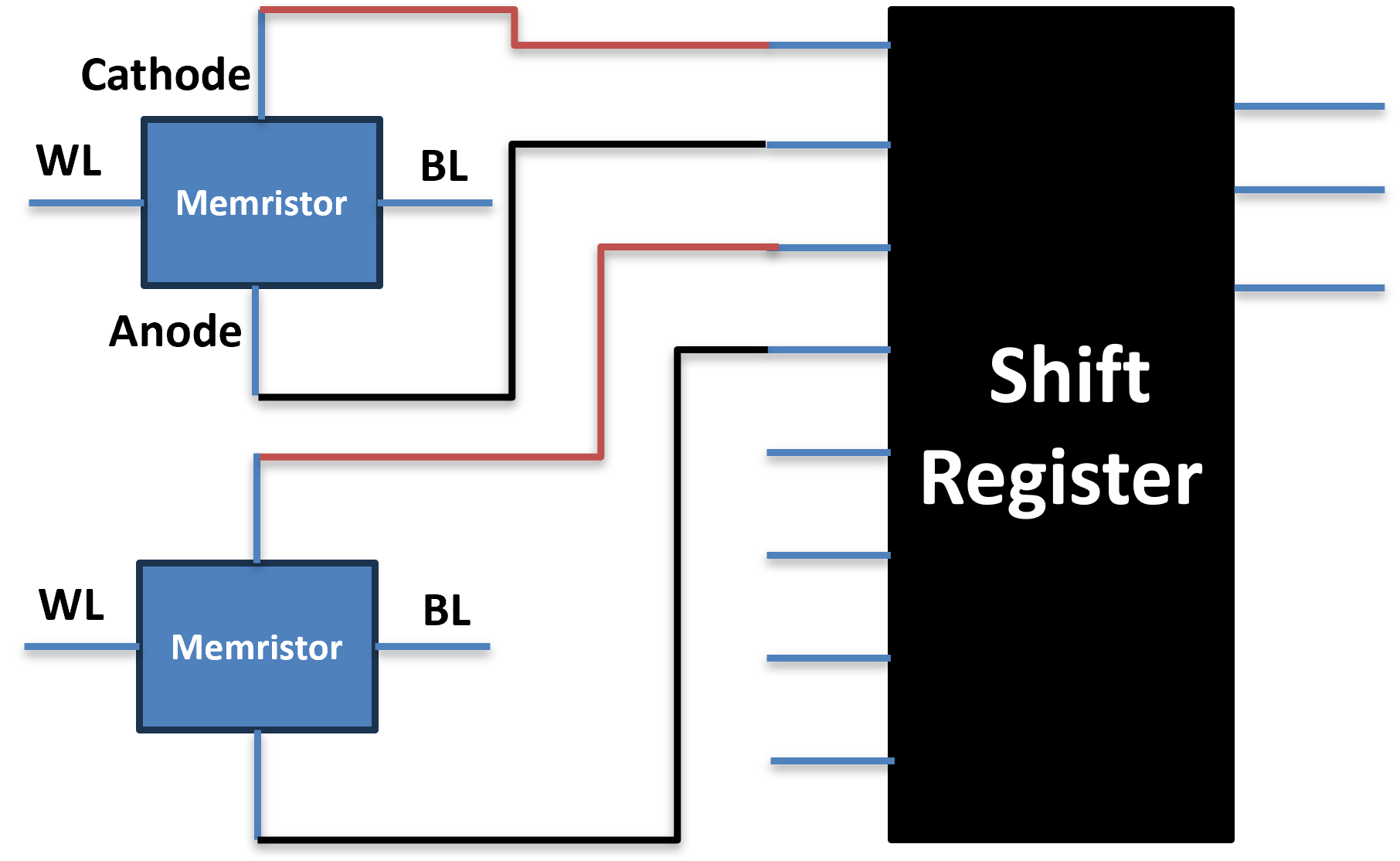}
    \caption{Proposed structure.}
    \label{fig:neo-struct}
\end{figure}

Our proposed architecture has two orthogonal paths, one path for read-only and another path for write-only.
One electrode pair maintains the conventional crossbar layout for data readout, while an orthogonal pair provides a dedicated write control path. Each cell is assigned an independent programming loop, eliminating unintended current paths during write operations. Consequently, cross-talk is not just mitigated but structurally eliminated. This design integrates seamlessly with shift registers and microcontroller-based control logic, as shown in Fig.~\ref{fig:neo-struct}.
The device operates by modulating a bulk material property, specifically, reversible control of carrier concentration throughout the active medium. From a first-principles standpoint, a memristive system must be capable of three core operations: a) integrate historical charge, functioning like an electrical integrator, b) modulate conductivity as a function of accumulated charge, and c) support reversible conductivity changes for reconfigurability.

Beyond the functional requirements of a memristive system, read–write separation further motivates the use of bulk-mediated resistance modulation as the functional carrier. Because bulk resistance reflects a global material property that can be accessed from multiple directions, resistance states programmed along the write path remain observable from an orthogonal read direction without engaging the write operation, allowing readout of the same underlying material state.

These criteria point to a reversible doping mechanism driven by external electric fields.
We achieve this via two cooperative physical processes. First, directional ionic motion enables controlled insertion of mobile ions into the device. Second, a polymeric host matrix serves as the charge storage medium, where specific functional groups (e.g., carbonyl, ether, nitrile) form coordination complexes with the inserted ions.

This mechanism is well-supported by solid-state battery research, which demonstrates that polymer electrolytes can reliably function as ion conductors and can host reversible ion movement within the polymer matrix \cite{yu2023battery,jacobs2022high}. Field-driven ion intercalation together with polymer-based charge storage form a complete physical framework that satisfies the orthogonal control requirements.

\begin{figure}[h]
    \centering
    \includegraphics[width=0.95\columnwidth]{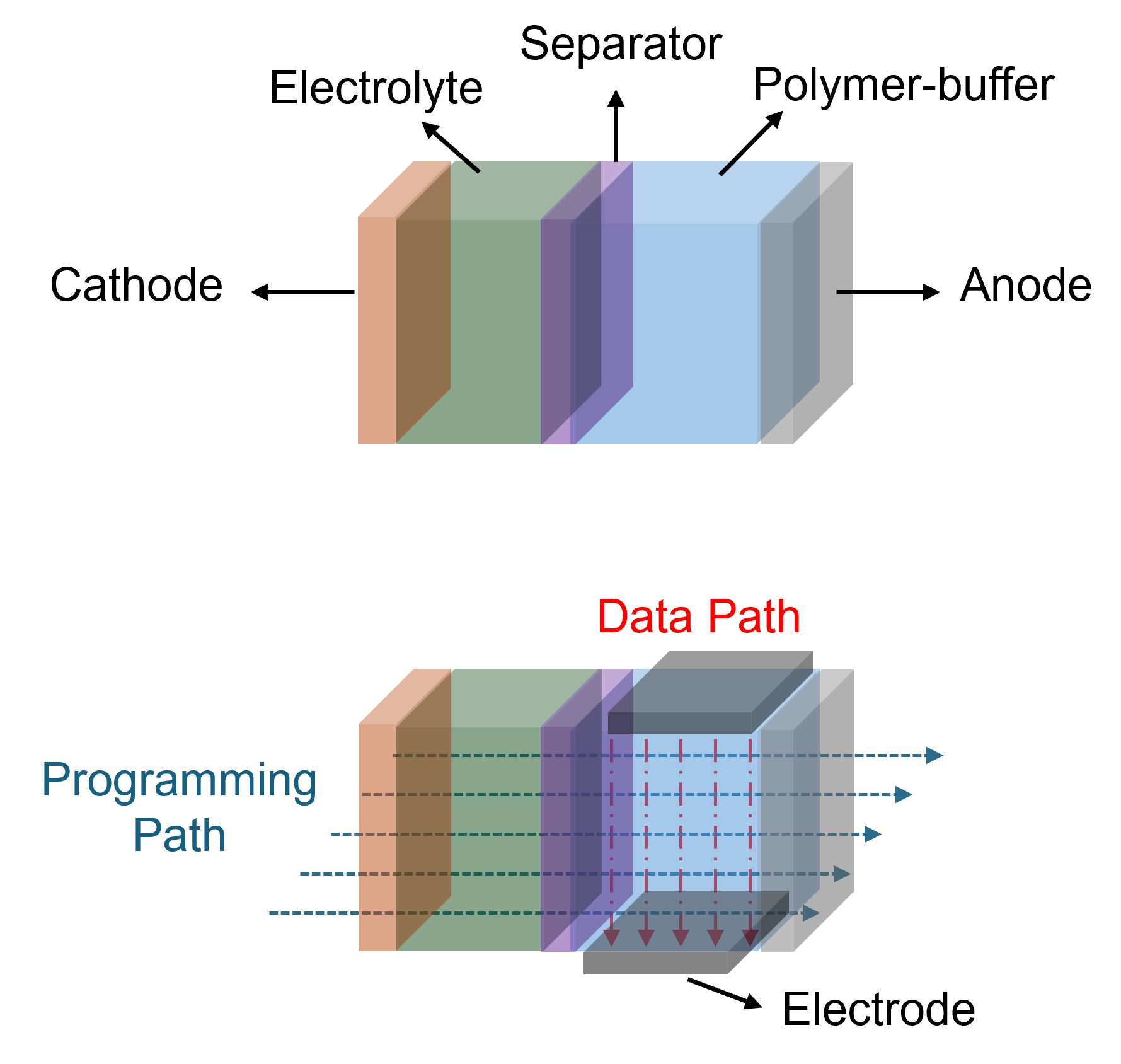}
    \caption{Structure of the memristor designed.}
    \label{fig:li-ion-structure}
\end{figure}

This concept is physically realized in the device structure shown in Fig.~\ref{fig:li-ion-structure}, where directional ionic programming and perpendicular electrical readout are implemented using a layered system with an ion acceptor layer.

\subsection{Electrical Properties of the Memristor Designed}\label{model}

\begin{figure}[h]
    \centering
    \includegraphics[width=0.95\columnwidth]{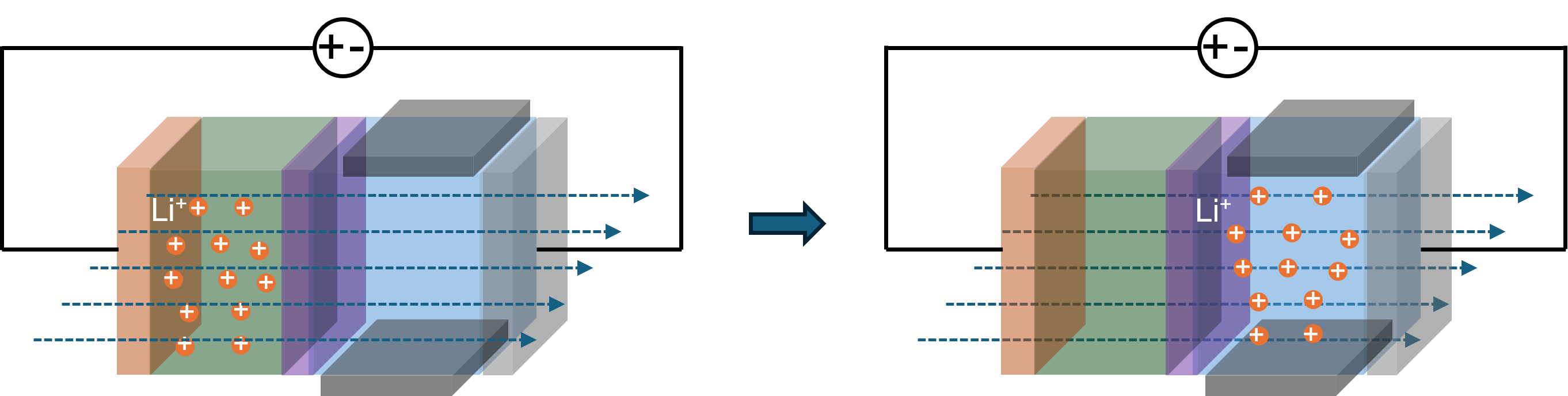}
    \caption{Principle of the proposed structure.}
    \label{fig:programming}
\end{figure}

During the write cycle, lithium ions in the electrolyte are driven by an electric field into the ion acceptor layer (Fig.~\ref{fig:programming}). This intercalation increases the ion acceptor layer's conductivity by introducing additional charge carriers, thereby reducing its resistivity. During discharge, the reversed field extracts lithium ions from the ion acceptor layer, increasing its resistivity. By varying the duration of charging and discharging, the lithium ion concentration (thus the resistance) can be fine-tuned. This enables continuous, reversible resistance programming, allowing the device to function as a memristor with adjustable states.

In Chua's memristor formulation, the magnetic flux $\phi(t)$ is a function of the total transported charge $q(t)$, and the memristance is defined as its derivative \cite{chua2003memristor}: $M(q) = d\phi/dq$.
A programming voltage $V_p$ drives lithium ions into the ion acceptor layer, generating a vertical electric field $E = V_p / d$, where $d$ is the cathode-to-anode distance. The ions drift velocity is $v = \mu E = \mu V_p / d$ ($\mu$ is the ionic mobility). Assuming a uniform initial ion concentration $c_0$ in the electrolyte, the resulting ionic current density magnitude is
\begin{equation}
J = e c_0 \mu V_p/d,
\end{equation}
where $e$ is the electric charge of an electron.
In the subsequent equation, the vector form $\vec{J}$  is used explicitly for the surface integral.
Integrating $\vec{J}$ over the cross-sectional area $A$ yields
the programming current expressed as
\begin{equation}
I_p = \iint_S \vec{J} \cdot d\vec{S} = e c_0 \mu \frac{V_p}{d} \cdot A
\end{equation}
where \(A=l_x l_z\) is the area of the cross section, \(l_z\) is the height of the readout electrode, $l_x$ is the distance between readout electrodes, and \(l_y\) is the thickness of the ion acceptor layer.

The total charge injected into the ion acceptor layer over time $t$ is $q(t) = I_p  t$.
These ions accumulate within the ion acceptor volume $A l_y$
establishing a local concentration
\begin{equation}
c(t) = q(t)/(e A l_y).
\end{equation}
The ion acceptor layer's conductivity linearly depends on $c(t)$:
\begin{equation}
\sigma(t) = \mu_e e c(t) = \mu_e q(t)/(A l_y)
\end{equation}
where $\mu_e$ is the effective electronic mobility within the ion-acceptor matrix, representing the average drift mobility of charge carriers generated by ion intercalation. This parameter governs the final conductivity of the doped region under an applied readout field.
The corresponding resistance is expressed as
\begin{equation}\label{EQ:7}
R(t) = l_x/\left( A \sigma(t) \right) = l_xl_y / \left( \mu_e q(t) \right).
\end{equation}
The read voltage becomes $V(t) = R(t)  I_r = l_xl_y  I_r/(\mu_e  q(t))$.
Integrating $V(t)$ over time yields the total magnetic flux:
\begin{equation}\label{EQ:9}
\phi(t) = \int_0^t V(\tau) \, d\tau = \int_0^t \frac{l_xl_y I_r}{\mu_e q(\tau)} \, d\tau = \frac{l_xl_y I_r}{\mu_e  I_p}  \ln t + C
\end{equation}
where we have used the relationship $q(t) = I_pt$.
Substituting $t = q/I_p$ into Eq. (\ref{EQ:9}) yields
\begin{equation}
\phi(q) = (l_xl_y I_r)/(\mu_e I_p) \ln\left( q/I_p \right) + C = K \ln q + C'
\end{equation}
where $K = (l_xl_y I_r)/(\mu_e I_p)$ and $C'$ absorbs the constant offset term $-(l_xl_y I_r)(\mu_e I_p)  \ln I_p$.

The memristance is thus obtained as
\begin{equation}
M(q) = d\phi/dq = K/q. \label{ktmodel}
\end{equation}
This result confirms that the proposed device inherently satisfies the memristor condition, exhibiting a continuous and explicitly derivable flux–charge relationship. Importantly, this behavior stems not from stochastic defect migration but from a well-defined, field-driven ionic doping mechanism.

Additionally, the analytical expression for $M(q)$ also reveals important properties of the system's dynamic programmability. The memristance as a function of time is
\begin{subequations}
\begin{align}
M(t) &= K/q(t) = K / (I_p t) \\
dM(t)/dt &= - K/(I_p t^2).
\end{align}
\end{subequations}

This function is continuous, differentiable, and monotonically decreasing with its magnitude diminishes over time. This indicates that the resistance changes become progressively more stable as more ions are injected. Practically, this enables analog programmability, allowing for arbitrarily fine resistance adjustments through precise control of programming duration.

Unlike conventional filament-based memristors that exhibit abrupt or binary resistance switching, the proposed structure enables smooth, deterministic modulation of conductivity. This behavior is well-suited for analog computing, neuromorphic systems, and multi-level memory, where fine resolution and repeatable tuning are critical.

In the theoretical model, the read current $I_r$ was treated as constant to enable a closed-form flux–charge relationship. In practice, however, $I_r$ is not externally controlled, since read operations are voltage-driven, and the current depends on the instantaneous resistance state. While this simplification enables analytical clarity, real-world control must rely on voltage–time profiles and system-level feedback mechanisms.

\subsection{Orthogonal Read and Write Paths via Spatial Separation}
Here we discuss the dynamics of the proposed memristor in a crossbar network under both write and read operations.

\subsubsection{Write Operation}
\begin{figure}[h]
    \centering
    \includegraphics[width=0.95\columnwidth]{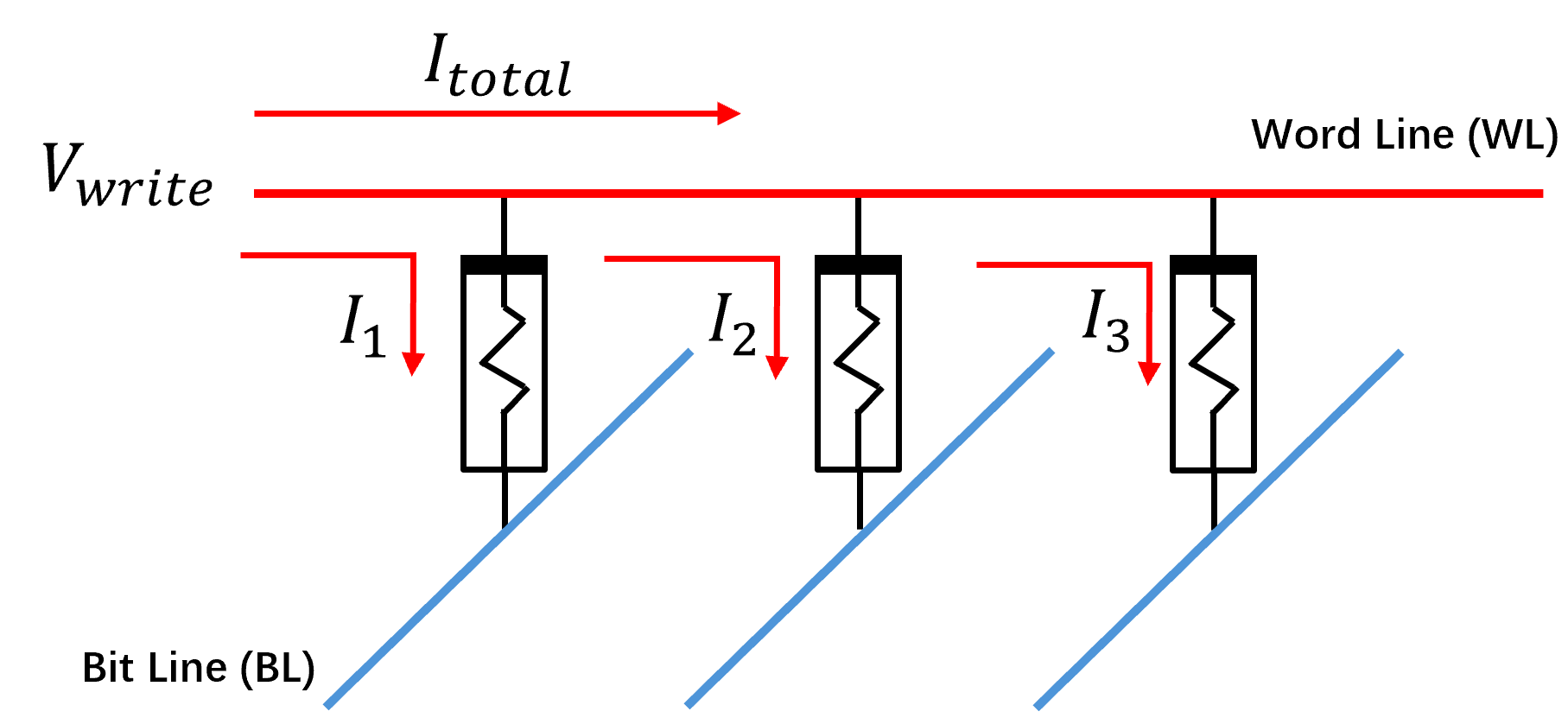}
    \caption{Conventional memristor and its crossbar network.}
    \label{fig:2tw}
\end{figure}

\begin{figure}
    \centering
    \includegraphics[width=0.95\columnwidth]{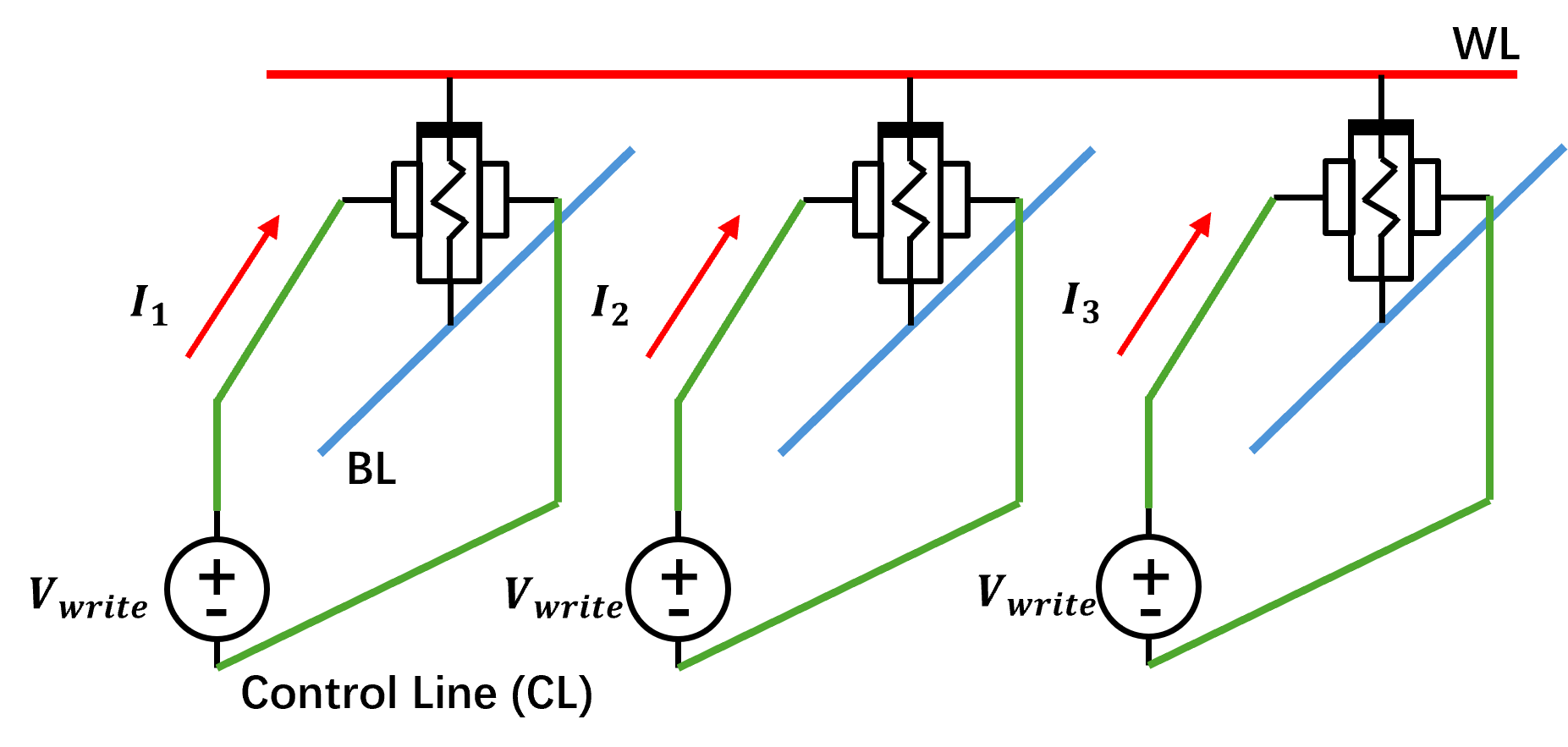}
    \caption{Proposed memristor and its crossbar network}
    \label{fig:4tw}
\end{figure}

Figs.~\ref{fig:2tw} and \ref{fig:4tw} compare the conventional and the proposed memristor network structures.

In conventional memristor networks, multiple cells share the same word line (WL), resulting in a partially parallel 3D crossbar topology where unintended current paths (sneak paths) are unavoidable. In the proposed structure (Fig. 5), each memristor has an independent control loop (CL$\pm$) for programming, which physically isolates its current from neighboring cells. This design eliminates sneak paths at the architectural level and enables fully parallel write operations.

\subsubsection{Read Operation}
The operation of the proposed structure during read is analyzed at both the network and device levels.

\begin{figure}[h]
    \centering
    \includegraphics[width=0.95\columnwidth]{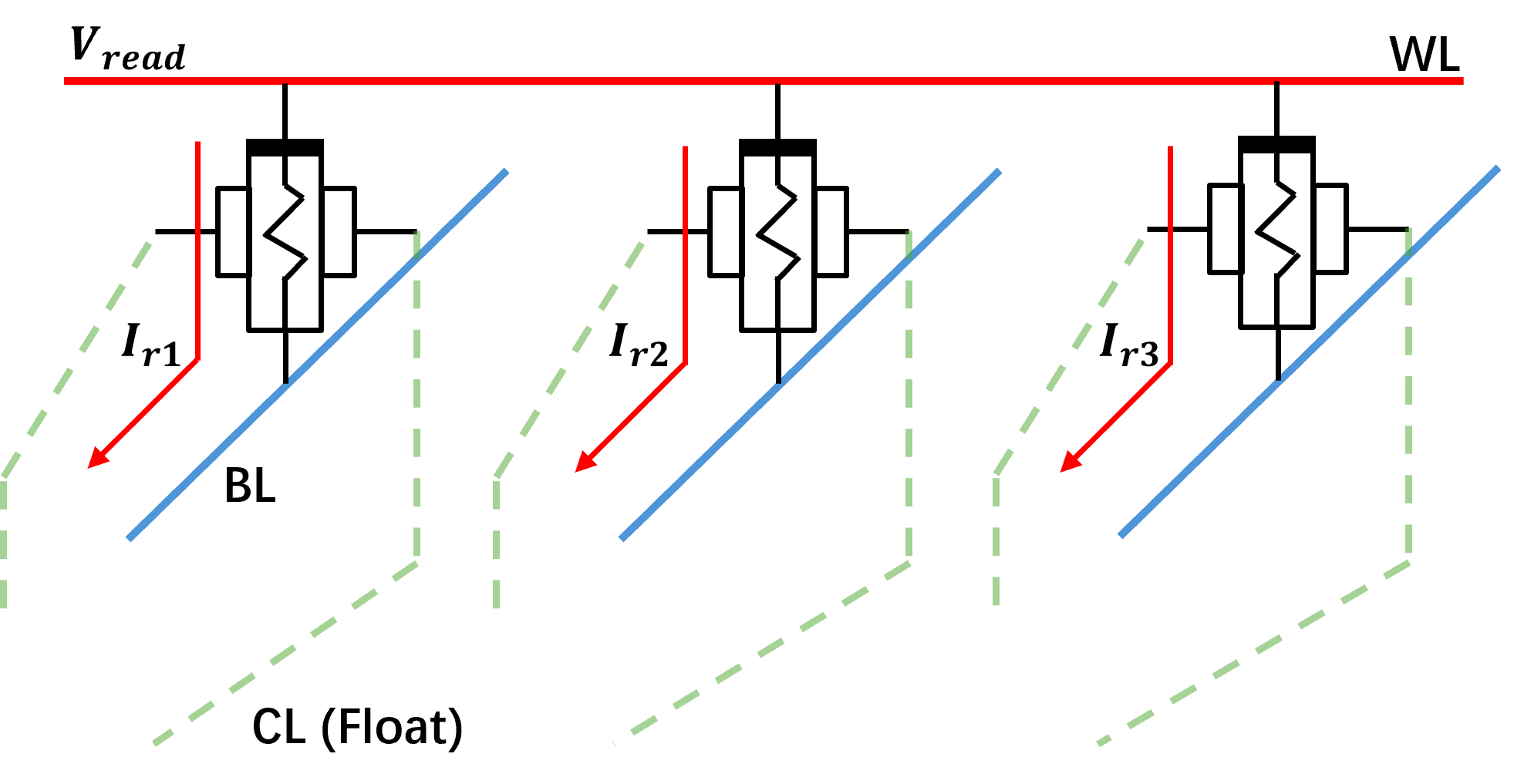}
    \caption{Structure during read operation at the network level.}
    \label{fig:read}
\end{figure}

At the network level (Fig.~\ref{fig:read}), the control lines CL$\pm$ are left floating during read. In this state, no current flows through the programming path, preventing any unintended write operation.

\begin{figure}[h]
    \centering
    \includegraphics[width=0.95\columnwidth]{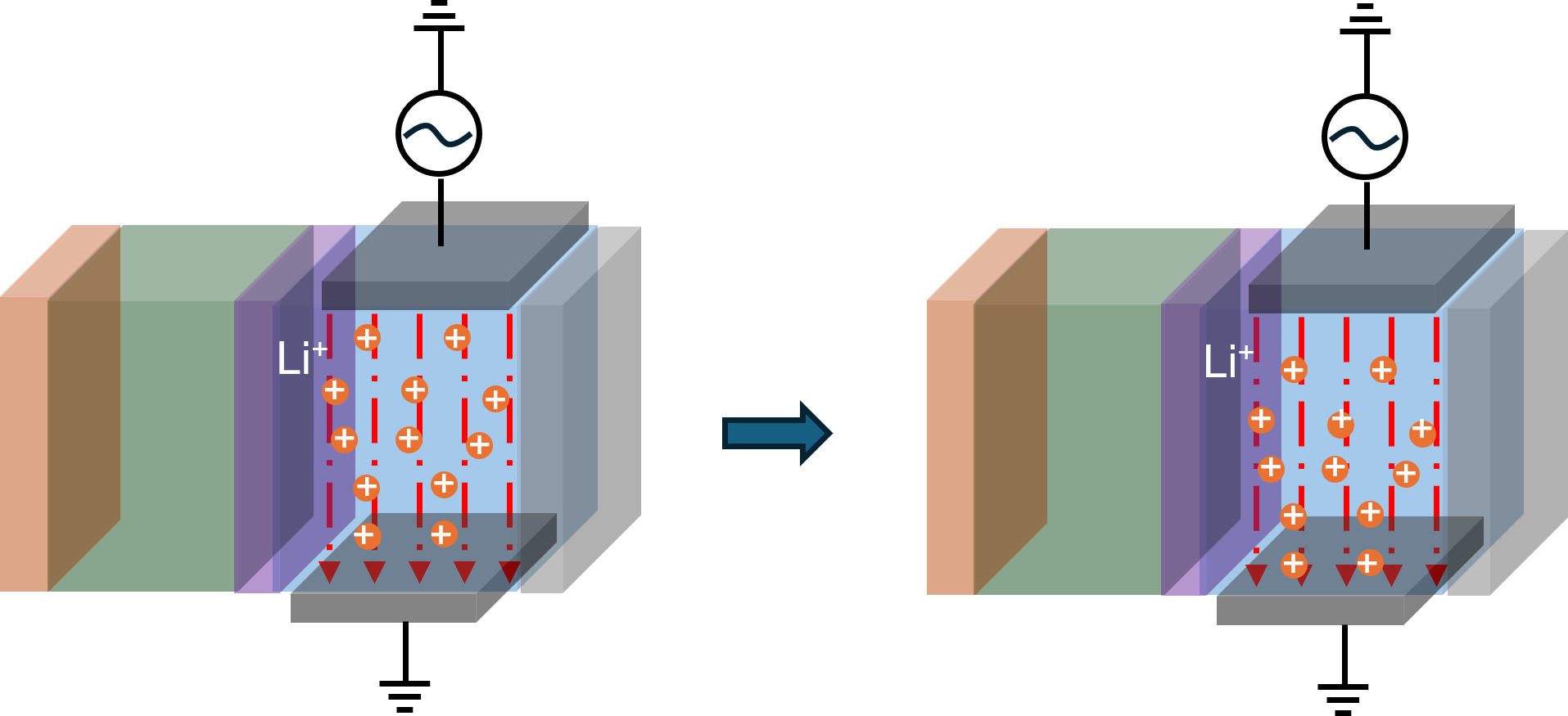}
    \caption{Structure during read operation at the device level.}
    \label{fig:read_device}
\end{figure}

At the device level, in the read mode the memristor only experiences an electric field along the read direction. Importantly, this electric field is orthogonal to the programming direction and therefore does not induce ion migration out of the ion acceptor layer.
As depicted in Figs.~\ref{fig:read_device} and \ref{fig:programming}, the orthogonality of read and write current paths ensures that the read process does not disturb the programmed conductance state. This design feature significantly mitigates the risk of long-term conductance drift triggered by repeated read operations.

\section{Experiment}
We fabricated two prototype memristors (hereafter referred to as samples S1 and S2) to validate the theoretical framework. In this experiment, an ion acceptor layer was prepared using polyethylene glycol (PEG), while the electrolyte consisted of \ce{Li+} based solution.
The electrochemical cell was assembled using a graphite anode, a \ce{LiFePO4} cathode, and a pair of aluminum readout electrodes.
This design enables reversible lithium-ion intercalation into the polymer matrix, thereby allowing continuous and tunable modulation of the device’s resistance states. During the experiment, the program voltage $V_p$ was set to $3.6\,\mathrm{V}$ to effectively stimulate lithium extraction.

\subsection{Programming Response}
\begin{figure}
    \centering
    \includegraphics[width=0.95\columnwidth]{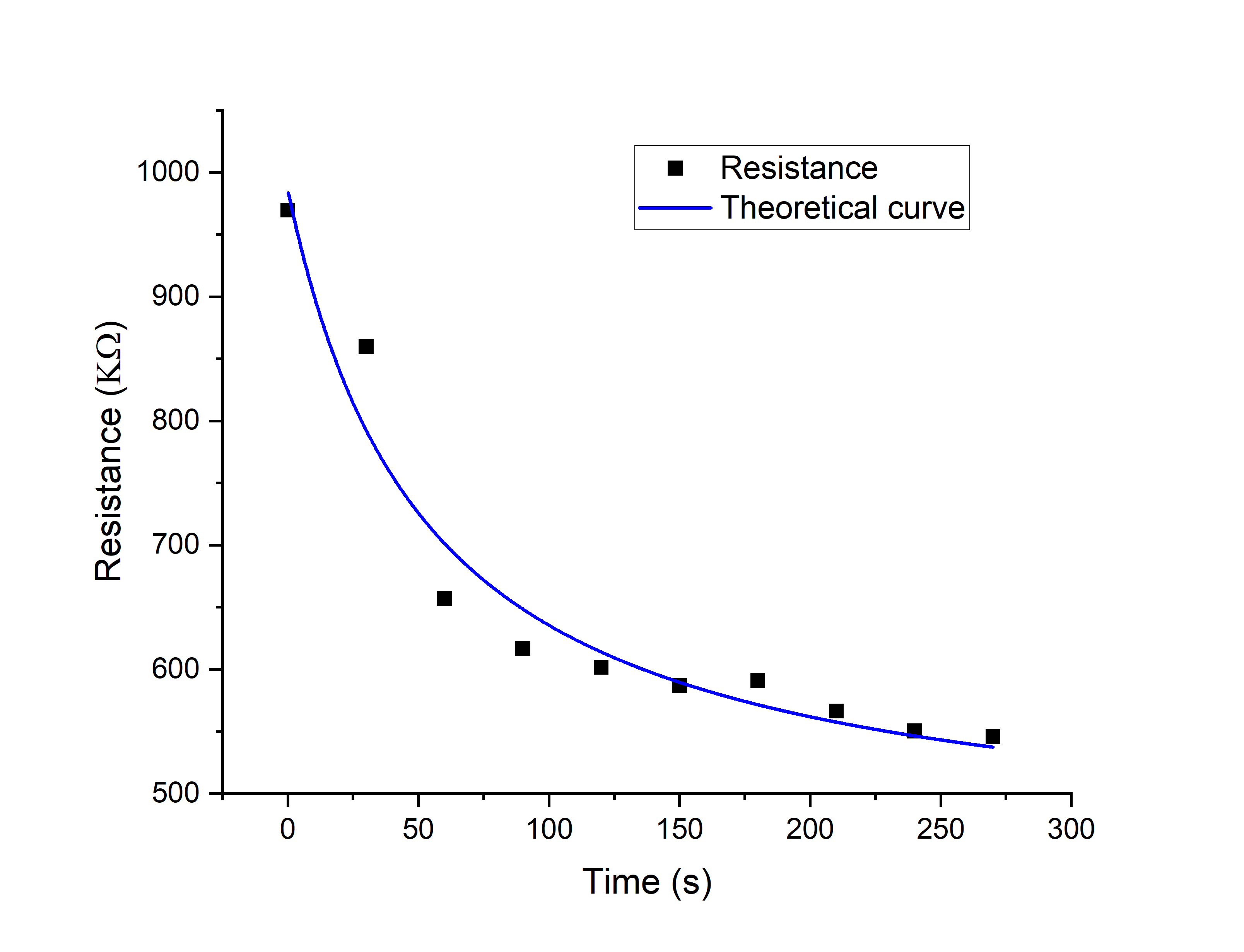}
     \caption{Measured resistance of sample S1 as a function of programming time and the theoretical curve from the model ($V_p = 3.6\,\mathrm{V}$).}
    \label{fig:tvsR1}
\end{figure}

The programming response of the fabricated device was characterized by programming sample S1 under $V_p$ for 30s in each cycle, followed by resistance measurement.
Fig.~\ref{fig:tvsR1} shows that the device exhibits a dynamic resistance range from a high-resistance state (HRS) of approximately $1\,\mathrm{M}\Omega$ to a low-resistance state (LRS) of approximately $550\,\mathrm{K}\Omega$. The evolution of the device resistance closely followed the trend predicted by Eq.~\eqref{ktmodel}, showing an initially rapid decrease followed by a gradual saturation behavior with increasing programming time. This behavior is consistent with the expected linear dependence of the ionic intercalation amount on programming time in the early stage and the saturation of available intercalation sites in the later stage.

\subsection{Programming, Retention, and Reverse-Programming} \label{fullfunction}
\begin{figure}[h]
    \centering

    \includegraphics[width=0.95\columnwidth]{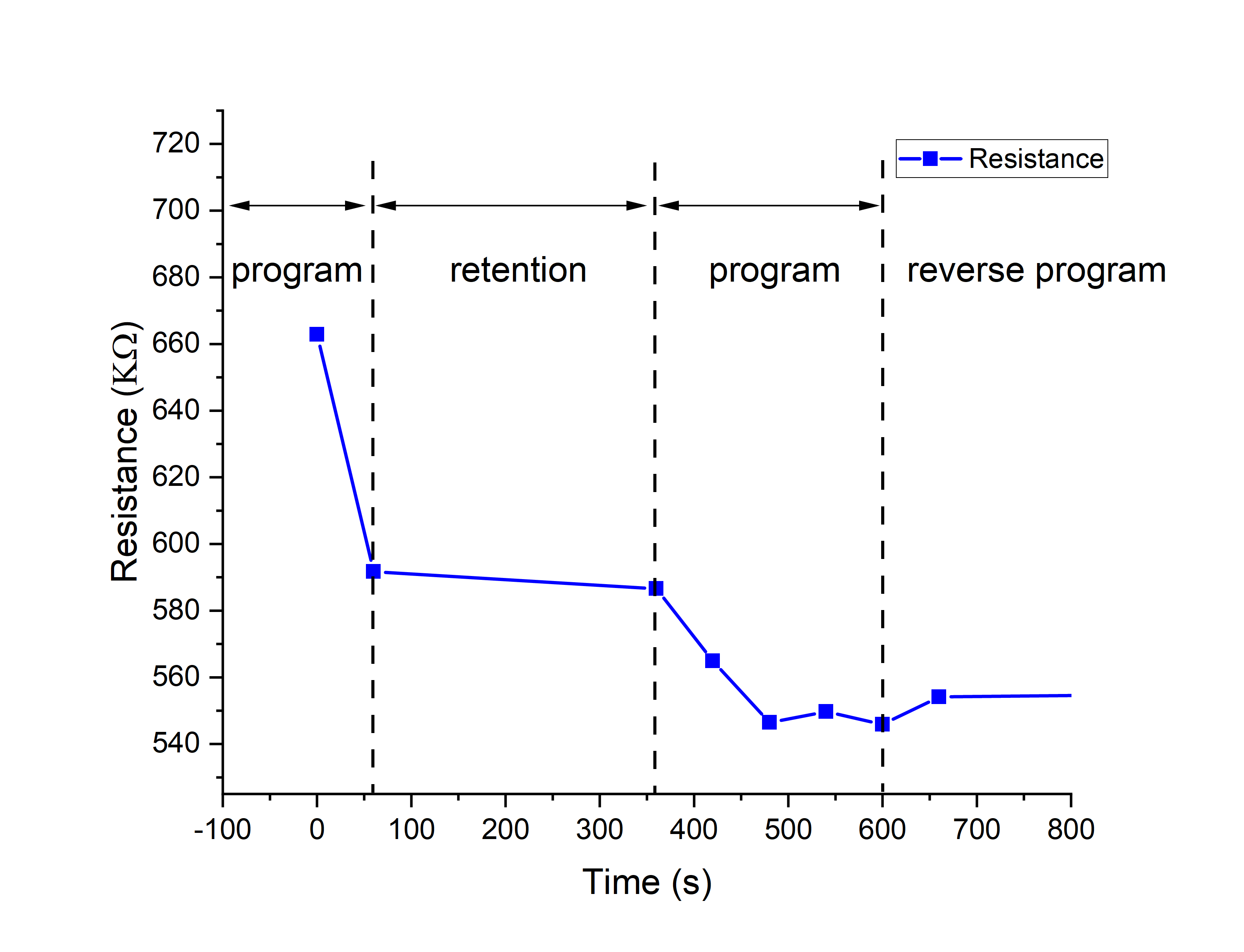}

    \caption{Dynamic resistance response of the memristor under different $V_p$, illustrating programming, retention, and reverse programming phases.}
    \label{fig:tvsR}
\end{figure}

In this experiment, the dynamic response of the fabricated device was evaluated under different $V_p$ conditions to validate its programming, reverse-programming, and retention functions. As shown in
Fig.~\ref{fig:tvsR}, sample S2 was first programmed at $V_p = 3.6\,\mathrm{V}$ for 60s, which brought its resistance down from the HRS. It was then kept at $0\,\mathrm{V}$ for 300s to verify the memory retention capability. The resistance from $t = 60\,\mathrm{s}$ to $t = 360\,\mathrm{s}$ remained stable with minimal drift, confirming that the programmed state could be maintained without bias.

Following the retention test, the device was reprogrammed at $V_p = 3.6\,\mathrm{V}$ for 60s intervals until saturation was observed. Finally, at $t = 600\,\mathrm{s}$, the device was reverse-programmed under $V_p = -3.6\,\mathrm{V}$ for 60s, which successfully increased the resistance, demonstrating that the device is also responsive to reverse polarity stimuli.

\subsection{Long-term Retention Characterization}
\begin{figure}
    \centering
    \includegraphics[width=0.95\columnwidth]{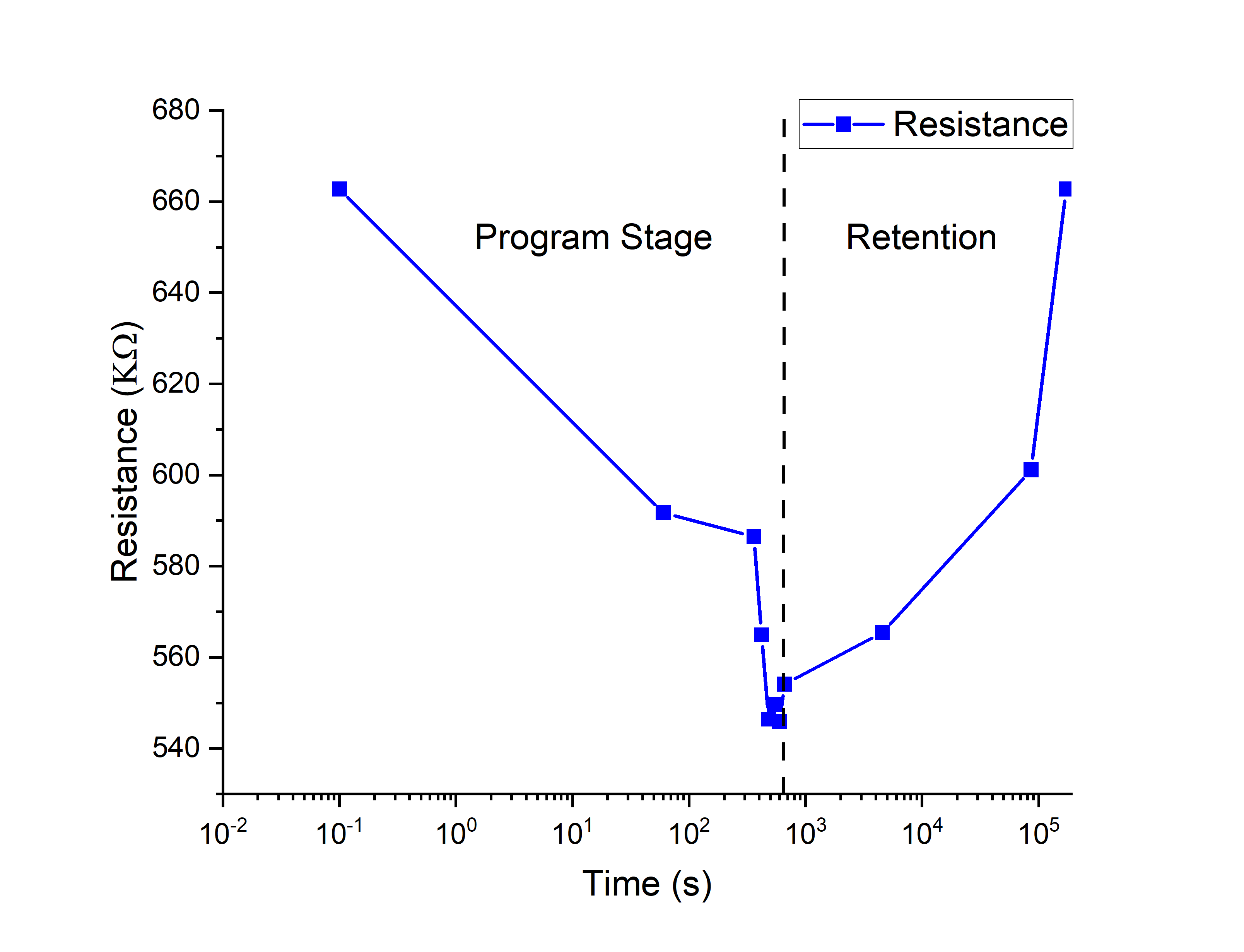}
    \caption{Time-dependent resistance relaxation after programming.}
    \label{fig:retention}
\end{figure}

Fig.~\ref{fig:retention} shows the time-dependent resistance relaxation of device S2 after programming. Under zero-bias conditions, the resistance gradually increases and returns close to its initial value within approximately 48 hours.

This relaxation behavior is primarily determined by the material system used in the present prototype. The ion acceptor layer is based on polyethylene glycol (PEG), where lithium ions are stored through relatively weak coordination interactions. As a result, the ionic potential wells are shallow, and thermal activation enables slow back-diffusion of ions once the external programming field is removed. This leads to a gradual reduction in carrier concentration and a corresponding resistance increase over long time scales.

The resistance evolution is fully reversible, and the device remains reprogrammable after relaxation. This behavior is consistent with a bulk ionic redistribution mechanism rather than irreversible filament formation.

\section{Discussion}
\subsection{On the Necessity of Array-Level Sneak-Path Validation}
Sneak-path formation in memristor crossbar arrays is a circuit-level consequence rather than an intrinsic device phenomenon. A necessary condition for sneak paths is the existence of unintended conductive paths that connect a selected cell to neighboring cells during programming or read operations.

In the proposed architecture, this condition is explicitly not satisfied. Each memristor cell forms an independent local programming loop that is physically and electrically isolated from all neighboring cells. At no operating mode does a conductive path exist that connects one cell to another. Consequently, sneak-path currents are precluded by construction rather than mitigated by biasing schemes or control protocols.

Under this topology, the absence of sneak paths constitutes a universal logical negation derived from circuit connectivity, not an empirical property to be exhaustively verified through array-level measurements. Array-level sneak-path experiments would therefore reduce to a consistency check of the predefined circuit constraints, rather than providing independent validation. Accordingly, their omission reflects the logical irrelevance of such measurements to the proposed architecture, not a lack of experimental completeness.

We note that in practical implementations, non-idealities (e.g., parasitics, unintended coupling, or fabrication defects) may cause the realized topology to deviate from the ideal circuit assumption and can introduce additional failure modes. A systematic array-level characterization of such non-ideal effects is beyond the scope of this paper .

\subsection{On the Absence of Pinched I–V Hysteresis and Threshold Voltage}
The pinched I–V hysteresis loop commonly used to identify memristive behavior is a representational construct rather than a fundamental requirement. In conventional filamentary devices, the hysteresis loop emerges from abrupt transitions between two discrete conductance states, typically associated with the formation and rupture of conductive filaments. In this case, the two branches of the hysteresis loop correspond to Ohmic responses with different effective conductances, while the apparent set and reset voltages mark the transition points between these discrete states.

In contrast, the proposed device operates under a fundamentally different programming principle. The conductance state is not controlled by exceeding a critical voltage threshold, but by the time-integrated programming stimulus. As a result, no intrinsic set voltage exists by design. Instead of voltage-triggered state transitions, the conductance evolves continuously as a function of the accumulated programming history.

From an electrical perspective, the slope of the I–V relation corresponds to the instantaneous conductance of the device. In systems where only two stable conductance states exist, repeated voltage sweeps naturally produce two dominant I–V branches, giving rise to a pinched hysteresis loop. However, when conductance is continuously tunable rather than discretized, the device does not map onto two distinct I–V curves. Instead, the admissible electrical states form a continuous set.

Accordingly, the electrical behavior of the proposed device is more appropriately described as a conductance manifold rather than a single projected I–V loop. Under voltage sweep measurements, this manifold collapses onto overlapping trajectories in the I–V plane, obscuring the underlying state evolution and eliminating the visual appearance of a pinched hysteresis. The absence of a conventional hysteresis loop therefore reflects the continuous and deterministic nature of conductance modulation, rather than a lack of memristive behavior.

In this context, the use of I–V hysteresis as a necessary validation criterion is not applicable. The device satisfies the defining characteristic of a memristive system through its history-dependent conductance evolution, while deliberately avoiding voltage-threshold-driven switching and discrete state transitions. Consequently, the lack of a pinched hysteresis loop is an inherent outcome of the design objectives, not a deficiency of the device.
\subsection{Routing Overhead as an Inevitable Cost of Parallelism}
We explicitly acknowledge that the proposed architecture introduces significant routing overhead. This consequence is fully anticipated and is not regarded as a drawback of the design. Rather, it represents the minimal and unavoidable cost of achieving true parallel programmability under realistic system constraints.

Parallelism, at its core, is the spatial redistribution of operations that would otherwise occur sequentially along the time axis. Any architecture that performs a set of write operations concurrently must externalize this concurrency into physical resources. In this sense, routing overhead is not an implementation artifact but a fundamental requirement of parallel execution. Eliminating temporal serialization necessarily demands proportional investment in spatial infrastructure.

Conventional crossbar architectures minimize routing by sharing access lines, thereby compressing physical resources at the expense of temporal scalability. This compactness enforces write contention, induces sneak-path constraints, and ultimately requires sequential or time-multiplexed programming. The proposed architecture makes the opposite and deliberate choice. By assigning an isolated programming loop to each cell, it removes structural coupling between write operations and enables deterministic full-parallel write.

This design choice inevitably incurs increased routing density and associated fabrication or packaging cost. These costs are not side effects to be optimized away, but prerequisites for eliminating write serialization. Any approach claiming full-parallel programmability without proportional spatial overhead merely shifts the cost into time-domain scheduling, selector complexity, or device-level non-idealities.

From a system-level perspective, this trade-off is rational and unavoidable. Time is a hard constraint that cannot be purchased once the architecture is fixed, whereas area, interconnect layers, and packaging complexity are soft constraints that can be addressed through engineering investment. The proposed architecture therefore prioritizes temporal scalability over spatial compactness, explicitly exchanging space for time.

Accordingly, routing overhead should not be interpreted as a limitation, but as the fundamental price of deterministic parallel execution. Evaluating the proposed approach under the assumptions of shared-line crossbar architectures is therefore conceptually incompatible with its design objectives. A detailed layout-level cost optimization is intentionally omitted, as footprint minimization is not the goal of this work. The contribution lies in establishing that architectural isolation is the necessary condition for full-parallel write, and that its physical cost is both explicit and irreducible.

\subsection{Impact of Device Geometry on Programming Speed}
The relatively slow programming speed observed in the present prototype should be interpreted in the context of its intentionally oversized physical dimensions. The proof-of-concept device reported in this work has a characteristic programming length scale on the order of 2 cm, which is several orders of magnitude larger than that of state-of-the-art nanoscale memristors.

In the proposed device, resistance modulation is governed by a field-driven ion intercalation process. Under this mechanism, the characteristic programming time is dominated by ionic drift rather than electronic transport. The ion drift velocity is given by \(v=\mu Vp/d\), where $\mu$ is the ionic mobility. Vp is the programming voltage, and d is the effective ion transport distance. As a result, the time required for ions to traverse the active region scales linearly with d.

For nanoscale filamentary or interfacial memristors, d typically ranges from a few nanometers to tens of nanometers. In contrast, the present prototype deliberately employs a macroscopic transport distance on the order of centimeters to simplify fabrication, enable direct probing, and validate the proposed physical principle. This difference in length scale alone introduces a slowdown exceeding ten orders of magnitude when compared to nanoscale devices, even under comparable electric fields.

Importantly, this reduced programming speed is not an intrinsic limitation of the ion-intercalation-based mechanism itself. Rather, it is a direct and predictable consequence of the macroscopic geometry chosen for this proof-of-concept implementation. The analytical model developed in Section~\ref{model} captures this dependence through the geometric parameters, and the experimental results are consistent with a drift-dominated transport regime.

When the same device structure is scaled down to the nanometer regime through materials engineering, the programming time can be substantially reduced. By shrinking the ion transport distance via nanoscale integration of the reservoir, electrolyte, and acceptor layers, the characteristic drift length d is reduced by several orders of magnitude, directly accelerating the ion-intercalation process under the same electric field. Importantly, such scaling does not require any change to the underlying operating principle of the device, but rather relies on established thin-film deposition, solid-state electrolyte engineering, and interface optimization techniques. Therefore, the present prototype should be understood as a macroscopic validation of the physical mechanism and architectural concept, while practical high-speed operation is expected to emerge naturally as the structure is scaled to nanometer dimensions through material and process optimization.
\section{Future Work}
While this work focuses on validating the physical mechanism and architectural feasibility of a memristor with orthogonal read and write paths, several directions remain for further development toward practical deployment.

First, future efforts will focus on identifying fully solid-state material systems through materials engineering. Replacing liquid or gel-based components with solid electrolytes and solid ion reservoirs is expected to improve device robustness, compatibility with microfabrication processes, and long-term stability. Solid-state implementations would also enable tighter geometric control of ion transport paths, which is essential for large-scale integration.

Second, retention performance can be further improved by engineering deeper ionic potential wells within the acceptor layer. By selecting materials and interfaces with stronger ion binding energies or higher activation barriers for back-diffusion, spontaneous relaxation of the programmed state can be suppressed. Such approaches are well established in solid-state electrochemistry and are directly applicable to the proposed device structure without altering its operating principle.

Third, device miniaturization to the nanometer scale is expected to significantly reduce programming time. In the present prototype, programming dynamics are governed by drift-dominated ion transport across a macroscopic distance. Scaling the structure down to nanometer dimensions through material and process optimization will proportionally reduce the ion transport length. Based on linear extrapolation from the drift model and the design parameters of the MEMS-compatible implementation, programming latency is expected to decrease by multiple orders of magnitude while maintaining the same voltage range and control methodology.

Finally, large-scale fabrication and statistical characterization will be necessary to evaluate device-to-device variability and programming consistency. Batch fabrication using MEMS or CMOS-compatible processes will enable systematic assessment of uniformity, yield, and endurance across large arrays. Such studies are critical for validating the scalability claims of the proposed architecture and for establishing confidence in its applicability to parallel in-memory computing systems.
\section{Conclusion}
This work establishes device-level decoupling of the shared read/write pathway as an effective strategy to preserve crossbar MAC functionality while eliminating sneak paths. Building on this structural principle, we developed a memristor that exploits controllable and reversible ion intercalation for stable, continuous resistance modulation. Preliminary results show near-ideal memristive behavior and good stability under isolated read/write operation. Continued material and interface optimization are expected to further enhance retention, stability, and drastically reduced programming time, underscoring the promise of this approach for reliable and scalable in-memory computing.

\bibliographystyle{IEEEtran}
\bibliography{ref}

\end{document}